\documentstyle[psfig]{article}

\def\be{\begin{equation}}
\def\bea{\begin{eqnarray}}
\def\ee{\end{equation}}
\def\eea{\end{eqnarray}}
\newcommand{\support}[1]{\thanks{#1}\ }

\def\fraction#1#2{{\textstyle{#1\over#2}}} 
\let\sfrac=\fr 

\def\pmb#1{\setbox0=\hbox{$#1$}%
  \kern-.025em\copy0\kern-\wd0
  \kern.05em\copy0\kern-\wd0
  \kern-.025em\raise.0433em\box0}
\def\journalfont{\it}         
\def\jou#1{{\journalfont #1\ }}
\def\PR{\jou{   Phys.\ Rev.}}
\def\CQG{\jou{  Class.\ Quantum Grav.}}
\def\vol#1{{\bf #1}}

\begin{document}
\title{Stationary Bianchi type II perfect fluid models}
\author{Ulf S Nilsson \\ 
       {\it Department of Physics, Stockholm University,}\\
       {\it Box 6730, S-113 85 Stockholm, Sweden} \\
        and \\
       Claes Uggla
       \support{Supported by the Swedish Natural Science Research Council}\\
       {\it Department of Physics, Stockholm University,}\\
       {\it Box 6730, S-113 85 Stockholm, Sweden} \\
       and \\
       {\it Department of Physics, Lule\aa\ University of
       Technology}\\
       {\it S--951 87 Lule\aa, Sweden} }
\maketitle

\begin{abstract}
  Einstein's field equations for stationary Bianchi type II models
  with a perfect fluid source are investigated. The field equations
  are rewritten as a system of autonomous first order differential
  equations. Dimensionless variables are subsequently introduced for
  which the reduced phase space is compact. The system is then studied
  qualitatively using the theory of dynamical systems. It is shown
  that the locally rotationally symmetric models are not
  asymptotically self-similar for small values of the independent
  variable. A new exact solution is also given. 
\end{abstract}

\centerline{\bigskip\noindent PACS number(s): 04.20.-q, 04.20.Jb, 04.40.Nr, 
95.30.Sf}
\newpage 

\section{Introduction}
\label{sec:intro}

Dynamical systems methods have been used for over 20 years for
studying the behavior of different models in general relativity,
especially in the field of cosmology, see e.g. \cite{dynsyst}. The
dynamical systems approach constitute a powerful tool when
one wants to study asymptotic and intermediate behavior.
It allows one to obtain a
good understanding of the models even though it may be impossible
to solve the corresponding equations exactly. 
Most of the attractors in the cosmological context have
turned out to be self-similar solutions but there are also more exotic
ones, e.g. the Mixmaster attractor. In this article we will apply the
dynamical systems approach to the stationary Bianchi type II models.
It will be shown that the locally rotationally symmetric (LRS)
models are not asymptotically self-similar for small values of the independent
variable. Instead the attractor is described by a heteroclinic cycle.
To our knowledge, the stationary LRS type II models yield the
simplest example of a non-self-similar attractor in general relativity. 

We will consider Bianchi type II models which admit a simply
transitive group of isometries acting on 3-dimensional hypersurfaces
which are timelike. The line element can be written as
$ds^2=\eta_{ab}\pmb{\omega}^a\pmb{\omega}^b$ where the
$\pmb{\omega}^a$'s are the 1-forms
\be
\pmb{\omega}^0 = D_1(x)\left(dt+cydz\right)\ ,\quad \pmb{\omega}^1=dx\ ,\quad
\pmb{\omega}^2 = D_2(x)dy\ ,\quad \pmb{\omega}^3 = D_3(x)dz\ ,
\ee
and $\eta_{ab}=\mathrm{diag}(-1,1,1,1)$.
The parameter $c$ is a constant. The source is assumed to be a perfect
fluid for which the energy-momentum tensor has the form $T_{ab} = \mu
u_au_b + p(\eta_{ab}+u_au_b)$ where $\mu$ is the energy-density, $p$ the
pressure, and $u^a$ the 4-velocity of the fluid. The
components of the fluid 4-velocity are $u^a=(1,0,0,0)$. An equation of
state of the form $p(\mu)=(\gamma-1)\mu$ with $1<\gamma<
2$ is also assumed. The metric coefficients $D_1,D_2$  
and $D_3$ are closely related to the kinematical properties of the
normal congruence of the symmetry surfaces. Note that this congruence
is spacelike. The expansion $\theta$ and the shear $\sigma_\pm$ are
given by
\bea
\theta &=& \frac{d}{dx}\left(\ln D_1D_2D_3\right)\ , \nonumber \\
\sigma_+ &=&-\frac{1}{2}\frac{d}{dx}\left(\ln
\frac{D_1^2}{D_2D_3}\right) \ , \quad 
\sigma_- = \frac{\sqrt{3}}{2}\frac{d}{dx}\left(\ln\frac{D_2}{D_3}
\right)\ .
\eea
We also define
\be
\label{ndef}
n = \sfrac{1}{4}c^2\left(\frac{D_1}{D_2D_3}\right)^2>0\ .
\ee
Einstein's equations, $G_{ab}=T_{ab}$, lead to
\paragraph{Evolution equations}
\bea
\dot{\theta} &=& - \sfrac13\theta^2 - \sfrac{2}{3}\sigma_+^2 -
\sfrac{2}{3}\sigma_-^2 - \sfrac12(2-\gamma)\mu \ ,\\
\dot{\sigma}_+ &=& -\theta\sigma_+ + 4n - \gamma\mu \ ,\\
\dot{\sigma}_- &=& -\theta\sigma_- \ ,\\
\dot{n} &=& -\sfrac{2}{3}(\theta+4\sigma_+)n\ ,
\eea
\paragraph{Defining equation for $\mu$}
\be
\label{mudef}
(\gamma-1)\mu = \sfrac13\theta^2 - \sfrac13\sigma_+^2 -
\sfrac13\sigma_-^2- n\ .
\ee
These equations can also be found from the orthonormal frame approach
by specializing the equations of \cite{NilssonUggla} to the present
models. The above set of equations is invariant under the
transformation $(\theta, \sigma_+, \sigma_-, n) \rightarrow (\theta,
\sigma_+, -\sigma_-, n)$. Therefore, without loss of generality, one can 
assume $\sigma_-\geq0$.  We can use the fact that $\mu$ is non-negative
together with equation (\ref{mudef}) to see that $\theta$ is a
``dominant'' quantity. Note also that $\theta$,
because of equation (\ref{mudef}), cannot change sign.

We now introduce $\theta$-normalized
variables:
\be
\label{thetanorm}
\Sigma_\pm = \frac{\sigma_\pm}{\theta}\ ,\quad N = \frac{3n}{\theta^2}
\ , \quad \Omega = \frac{3\mu}{\theta^2}\ .
\ee
The introduction of a dimensionless independent variable $\eta$
according to $\theta dx = 3d\eta$, leads to a decoupling of the
$\theta$-equation,
\be
\theta^\prime = \frac{d\theta}{d\eta} = -(1+q)\theta\ ,\quad  q :=
2\Sigma_+^2 + 2\Sigma_-^2+ \sfrac12(2-\gamma)\Omega \ . 
\ee
The remaining equations can now be written in dimensionless form:
\paragraph{Evolution equations}
\bea
\Sigma_+^\prime &=& -(2-q)\Sigma_+ + 4N - \gamma\Omega \ ,\\
\Sigma_-^\prime &=& -(2-q)\Sigma_- \ ,\\
N^\prime &=& 2(q-4\Sigma_+)N\ .
\eea
\paragraph{Defining equation for $\Omega$}
\be
(\gamma-1)\Omega = 1 - \Sigma_+^2 - \Sigma_-^2- N\ .
\ee
The boundary consists of a number of invariant sets which are
important in understanding the dynamics of interior orbits and we will
therefore include them. Moreover, this yields a compact reduced phase
space. The boundary is given by (i) the static 
Bianchi type I models, $N=0$, and (ii) the vacuum submanifold,
$\Omega=0$. There is also the locally rotationally symmetric (LRS) 
submanifold given
by $\Sigma_-=0$ which divides the phase space into two parts, related by 
the discrete symmetry $\Sigma_- \rightarrow -\Sigma_-$. Note that the 
rotation of the fluid is non-zero, since $\omega^2=n = N\theta^2/3$ 
\cite{NilssonUggla}.

\section{Dynamical systems analysis}
\label{sec:local}
We start by listing the equilibrium points and the corresponding value of
$\Omega$, which shows if the point is located on the vacuum
submanifold or not. The eigenvalues for each point are also given but
we refrain from giving the eigenvectors explicitly. 

\paragraph{The equilibrium points $K$}
\bea
& & \Sigma_+^2 + \Sigma_-^2 = 1 \ ,\, N = 0\ ;\, \Omega = 0 \\
& &  \frac{(5\gamma-6) + 2\gamma\Sigma_+}{\gamma-1}\ , \quad
4(1-2\Sigma_+)\ , \quad 0 \ .
\eea

\paragraph{The equilibrium point $W$}
\bea
& & \Sigma_+ = \frac{2-\gamma}{2(5\gamma-4)}\ ,\, 
\Sigma_- = 0\ ,\, N = \frac{3(5\gamma^2-4)}{4(5\gamma-4)^2}\ ;\,
\Omega = \frac{3(7\gamma-6)}{(5\gamma-4)^2} \\
& &  \frac{-12(\gamma-1)^2 \pm
  3\sqrt{\lambda}}{2(5\gamma-4)(\gamma-1)}\ ,\quad
-12\frac{\gamma-1}{5\gamma-4}\ ,
\eea
where
\be
\lambda = 2(\gamma-1)\left[8(\gamma-1)^3 - (7\gamma-6)(5\gamma^2-4)\right]\ .
\ee
The first two eigenvalues of $W$ are always complex with
a negative real part for $1<\gamma<2$.
The point lies in the LRS
submanifold and corresponds to the self-similar solution 
in \cite{wainwright_disc}.

Secondly we note that the equations corresponding to the static Bianchi type I
boundary, $N=0$, can be solved exactly. We find that
\be
\Sigma_- =
\frac{s_-}{1+As_+}\left(1+A\Sigma_+\right)\ ;
\quad A := \frac{5\gamma-6}{2\gamma}\ ,
\ee
where $s_\pm$ are constants satisfying $s_+{}^2 + s_-{}^2 = 1$.
The vacuum boundary, $\Omega=0$, is also solvable:
\be
\Sigma_- = \frac{s_-}{s_+ - 2}\left(\Sigma_+ - 2\right)\ ,
\quad N = 1-\Sigma_+^2 - \Sigma_-^2 \ .
\ee
A non-LRS exact solution which is characterized by a constant value of
$\Sigma_+=(2-\gamma)/(2(5\gamma-4))$ can also be found (the solution was found using the Hamiltonian approach developed in \cite{UgglaKT}. See the appendix
for the explicit line element). The orbit starts on the Kasner circle
$K$ and ends at the point $W$. This solution is important as the
orbits in the interior non-LRS part of the phase space spiral around it.
The situation is analogous to that of the spatially homogeneous Bianchi type
II models \cite{uggla}.

The system of differential equations
admits an increasing monotone function $Z$ in the interior phase space, 
excluding the point $W$ where $Z$ takes its maximum value 
(this monotone function has been found by using the Hamiltonian methods 
developed in chapter 10 in \cite{dynsyst}).
The function is given by
\be
Z = \frac{N^m\Omega^{1-m}}{(1-v\Sigma_+)^2}\ ,\quad v :=
\frac{2-\gamma}{2(5\gamma-4)}\ , \quad m :=
\frac{5\gamma^2-4}{(11\gamma-10)(3\gamma-2)}\ ,
\ee
with
\be
\frac{Z^\prime}{Z} = 16(\gamma-1)\frac{ \left(2(5\gamma-4)\Sigma_+ - (2-\gamma)\right)^2 + 3(3\gamma-2)(11\gamma-10)\Sigma_-^2}{(11\gamma-10)(3\gamma-2)\left[2(5\gamma-4)-(2-\gamma)\Sigma_+\right]}\ .
\ee
This monotone function prevents the existence of equilibrium points,
periodic orbits, recurrent orbits and homoclinic orbits in
this region, see e.g. \cite{Hsu}. 

The phaseportraits of the boundaries are given in figure 1a-d, while
the phaseportrait of the LRS submanifold is shown in figure 1e. We see
from the latter that orbits asymptotically approach
$W$ when $\eta\rightarrow\infty$, while for $\eta\rightarrow-\infty$
there exists a heteroclinic cycle, described by the LRS type I and the LRS
vacuum submanifolds. Figure 1f depicts the phase space
of the Bianchi type II non-LRS models. Note, in particular, the exact
solution characterized by $\Sigma_+=(2-\gamma)/(2(5\gamma-4))$.
All other non-LRS orbits start at the Kasner circle, $K$,
and spiral around this orbit towards $W$.
In this case the LRS heteroclinic cycle is no longer an attractor. Instead
it describes the intermediate behavior of those orbits which come ``close'' 
to it.

\begin{figure}
  \centerline{ \hbox{
      \psfig{figure=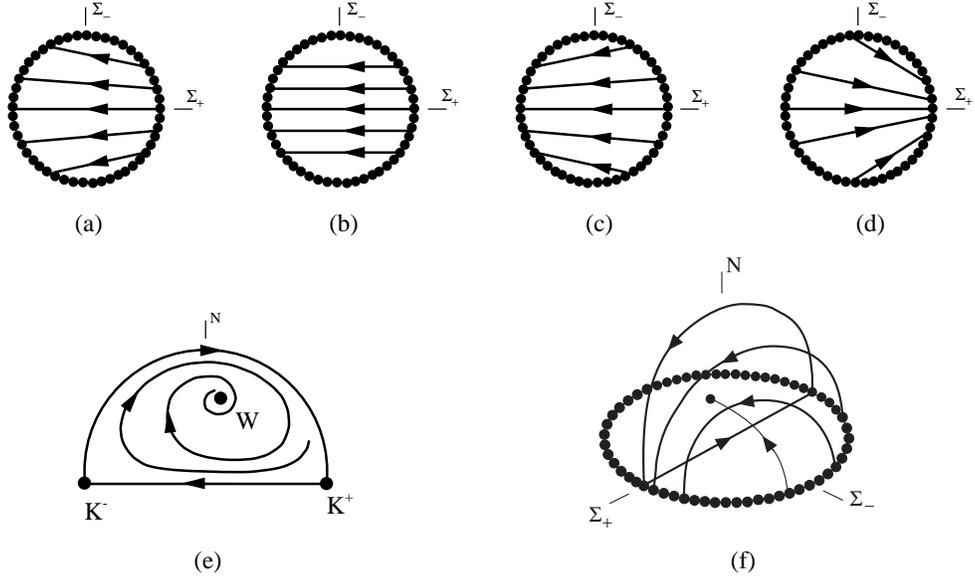}}}
  \caption{The phase portraits of the stationary Bianchi type
      II models corresponding to (a) the $N=0$ boundary for
      $1<\gamma<6/5$, (b) the $N=0$ boundary for $\gamma=6/5$, (c) the
      $N=0$ boundary for $6/5<\gamma<2$, (d) the vacuum boundary,
      $\Omega=0$ projected onto the $\Sigma_\pm$-plane, (e) the LRS 
      submanifold, $\Sigma_-=0$ and (e) the
      full phase space.}
    \label{fig:static}
\end{figure}

So far we have only discussed the mathematical features of the stationary
type II models. However, they might also be of some physical interest. 
Wainwright has speculated that the solution corresponding
to the equilibrium point $W$ might be interpreted as an approximation to the 
interior of a rotating disc of matter \cite{wainwright_disc}.
This interpretation should also pertain to the non-self-similar 
LRS models and perhaps also to the non-LRS models since they asymptotically
approach $W$. 
However, the main importance of the present models is probably
as part of a bigger picture where they may act as building blocks. 
The phase space of the present models form part of the boundary of more 
general stationary Bianchi models and hypersurface self-similar models
(see \cite{NilssonUggla}). These models in turn form part of the boundary
of more general models like the physically interesting $G_2$-models, see
e.g. \cite{WainwrightG2}. When studying these models one thus have to
be observant of the behavior associated with the present heteroclinic cycle
which could be expected to describe asymptotic or intermediate oscillating
spatial behavior.

\section*{Acknowledgments}

CU is supported by the Swedish Natural Science Research Council.

\appendix
\section{The non-LRS exact solution}
The line element for the non-LRS exact solution is given by
\bea
ds^2 &=& -[x(x+a)]^{1/2}\left( dt + cydz \right)^2 + [x(x+a)]^{q}dx^2
\nonumber \\
& & \quad + x^{p_+}(x+a)^{p_-}dy^2 + x^{p_-}(x+a)^{p_+}dz^2\ ,
\eea
with
\bea
q &=& -\frac{3\gamma-4}{4(\gamma-1)}\ ,\quad  p_\pm =
\frac{3\gamma-2 \pm \delta}{8(\gamma-1)} \ ,\nonumber \\
\delta &=& \sqrt{(11\gamma-10)(3\gamma-2)} \ ,\quad c^2 =
\frac{5\gamma^2 - 4}{16(\gamma-1)^2}\ ,
\eea
where $a$ is a constant. Setting $a=0$ yields the self-similar solution of
\cite{wainwright_disc}. For the remaining non-self-similar solutions one
can set $a=1$ by using the scale invariance.


\newpage

\begin{thebibliography}{00}
  \bibitem{dynsyst}
  J. Wainwright and G. F. R. Ellis ed., {\it Dynamical systems in
  cosmology} (Cambridge University Press, Cambridge, 1997)
  \bibitem{NilssonUggla}
  U. S. Nilsson and C. Uggla, Hypersurface homogeneous and
  hypersurface self-similar models, {\em submitted to
    Class. Quant. Grav}
\bibitem{wainwright_disc}
  J. Wainwright, {\em Galaxies, Axisymmetric systems and
  Relativity}, ed. M. A. H. MacCallum (Cambridge University Press, Cambridge, 
  1985) 
\bibitem{UgglaKT}
 C. Uggla, R. T. Jantzen and K. Rosquist, \PR D \vol{51}, 5522 (1995)
\bibitem{uggla}
  C. Uggla, \CQG \vol{6}, 383 (1989)
\bibitem{Hsu}
  J. Wainwright and L. Hsu, \CQG \vol{6}, 1409 (1989)
\bibitem{WainwrightG2}
  C. G. Hewitt and J. Wainwright, \CQG \vol{7}, 2295 (1990)
\end{thebibliography}
\end{document}